\documentclass[prl,preprint,showpacs]{revtex4}
\usepackage{graphicx}
\usepackage{color}
\usepackage{portland}

\begin{document}

%\usepackage{amssymb,amsfonts,amsmath}
%\definecolor{orange}{rgb}{1,0.5,0}
%\definecolor{purple}{rgb}{0.5,0,0.5}

%%%%%%%%%%%%%%%%%%%%%%%%%%%%%%

\title{Ultrafast structural and electronic dynamics of the metallic phase in a layered manganite.} 

\author{L. Piazza}
\affiliation{Laboratory for Ultrafast Microscopy and Electron Scattering, ICMP, Ecole Polytechnique F\'{e}d\'{e}rale de Lausanne, CH-1015 Lausanne, Switzerland}
\author{C. Ma}
\affiliation{Beijing National Laboratory for Condensed Matter Physics, Institute of Physics, Chinese Academy of Sciences, Beijing 100190, People’s Republic of China}
\author{H. X. Yang}
\affiliation{Beijing National Laboratory for Condensed Matter Physics, Institute of Physics, Chinese Academy of Sciences, Beijing 100190, People’s Republic of China}
\author{A. Mann}
\affiliation{Laboratory for Ultrafast Microscopy and Electron Scattering, ICMP, Ecole Polytechnique F\'{e}d\'{e}rale de Lausanne, CH-1015 Lausanne, Switzerland}
\author{Y. Zhu}
\affiliation{Department of Condensed Matter Physics, Brookhaven National Laboratory, Upton, New York 11973, USA}
\author{J. Q. Li}
\affiliation{Beijing National Laboratory for Condensed Matter Physics, Institute of Physics, Chinese Academy of Sciences, Beijing 100190, People’s Republic of China}
\author{F. Carbone}
\affiliation{Laboratory for Ultrafast Microscopy and Electron Scattering, ICMP, Ecole Polytechnique F\'{e}d\'{e}rale de Lausanne, CH-1015 Lausanne, Switzerland}

%%%%%%%%%%%%%%%%%%%%%%%%%%%%%%%%%%%%%%%%%%%%%%%%%%%%%%%%%%%%%%%%

\begin{abstract} 

The transition between different states in manganites can be driven by various external stimuli. Controlling these transitions with light opens the possibility to investigate the microscopic path through which they evolve. We performed femtosecond (fs) transmission electron microscopy on a bi-layered manganite to study its response to ultrafast photoexcitation. We show that a photoinduced temperature jump launches a pressure wave that provokes coherent oscillations of the lattice parameters, detected via ultrafast electron diffraction. Their impact on the electronic structure are monitored via ultrafast electron energy loss spectroscopy (EELS), revealing the dynamics of the different orbitals in response to specific structural distortions.

\end{abstract}

\maketitle

Layered materials provide a unique playground for charge ordering phenomena due to the lowered dimensionality of their electronic structure \cite{graphenereview, manganitereview, cuprates}. These states of matter are of current interest both for their fundamental properties and their potential exploitation in microelectronics. In particular, manganites have a rich phase diagram originating from the presence of several competing energy scales; this results in a broad versatility which has already seeded multiple applications \cite{manganitesappl}. 
Transitions between different magnetic, charge and orbital ordered states, can be driven by temperature \cite{manganitetemp}, pressure \cite{manganitepress}, magnetic field \cite{manganitemagfield}, and chemical doping \cite{manganitetemp}. Recent experiments have shown that light pulses can also make these systems cross the different critical lines of their phase diagram \cite{rini09}. In particular,  the dynamics of charge and orbital ordering have been investigated by directly triggering specific lattice distortions via resonant THz radiation pulses, \cite{cav07nm} and orbital waves have also been detected via high temporal resolution optical experiments \cite{cav07n}. Moreover, the temporal evolution of the net magnetization upon photoexcitation has been observed by ultrafast optical Kerr rotation experiments \cite{fskerr1, fskerr2}, yielding quantitative information about the spin-orbit interaction in La$_{0.6}$Sr$_{0.4}$MnO$_{3}$. So far, most of the time-resolved studies in these materials have been carried out via optical \cite{taylor01, fskerr3}, and X-ray probes \cite{cav11, cav11b, steve1, steve2, fsxray}; the latter have allowed the observation of the structural rearrangement in the solid upon melting the charge and orbital ordering \cite{steve1, steve2}, as well as the formation of metastable states with exotic properties accessible only in the picosecond (ps) time scale \cite{ichi11}. Combined dynamical information about the crystal and electronic structure have been obtained via resonant X-ray diffraction at the L absorption edge of Mn ions, very sensitive to the material's spin and valence state, showing that in the metastable photoinduced phase, magnetism could be suppressed while maintaining the orbital ordering in the system \cite{cav11b}. 
Charge and orbital organized patterns in solids give rise to spatial modulations of the crystalline order visible as superlattice peaks in diffraction, often accompanied by distinct fingerprints in the electronic structure. As demonstrated in \cite{cav11b}, the ability to monitor the evolution of both the crystal and electronic structure in a broad momentum and energy range provides a privileged point of view on these phenomena and their evolution through the critical lines of the phase diagram.

Charge and orbital ordering in strongly correlated solids have been observed also by electron diffraction and microscopy, both statically \cite{chao} and with fs resolution \cite{jure, baum, nuh}. 
Ultrafast electron microscopy offers the advantage of delivering momentum-resolved information through diffraction and broad-band electron energy loss spectroscopy in a very direct fashion and from very tiny amounts of material \cite{carbonescience, carbonecpl, carboneperspective}. In the bi-layered PrSr$_{0.2}$Ca$_{1.8}$Mn$_2$O$_7$, a peculiar checkerboard pattern made of ordered orbital stripes has been observed via transmission electron microscopy (TEM) \cite{chao}, and optical spectroscopy \cite{prmno1}, and originates from a 1:1 ratio of Mn$^{3+}$ and Mn$^{4+}$ ions in the ground state. The evolution of the orbital and charge ordering in this material can be tuned with chemical doping or temperature, giving rise to a very rich phase diagram \cite{phased}.

In this work, we show that laser light can be used in-situ in a TEM to drive  a PrSr$_{0.2}$Ca$_{1.8}$Mn$_2$O$_7$ sample in different regions of its phase diagram, and perform an ultrafast investigation of its metallic phase. Pressure waves can be launched in thin samples via light-induced temperature jumps, resulting in a modulation of the crystal structure that can be followed by ultrafast diffraction and imaging experiments \cite{brettscience, nanodrum}, while the consequent modulation of the electronic structure can be monitored with ultrafast spectroscopy \cite{carbonescience, carbonecpl}. Coherent structural motions of a PrSr$_{0.2}$Ca$_{1.8}$Mn$_2$O$_7$ single crystal induced by the photoinduced temperature jump were detected as periodic modulations of the Bragg peak intensities. In the same instrument, the resulting modulation of the loss function was recorded over a broad energy range (2 eV to 70 eV), evidencing the response of the different electronic states to specific atomic motions. Such an interplay was understood via density functional theory (DFT) calculations of the EELS spectra, performed considering an equilibrated electronic structure at  different lattice parameters.
This study provides the necessary background information to further investigate the more complex regions of the phase diagram, and confirms the ability of the ultrafast temperature-modulation approach to provide a direct view of light induced thermal phenomena in a solid.

% Experiments

A polycrystalline sample of PrSr$_{0.2}$Ca$_{1.8}$Mn$_2$O$_7$ was prepared by the solid-state reaction technique. The compound was prepared via a mixture of Pr$_6$O$_{11}$, CaCO$_3$, SrCO$_3$ and MnO$_2$, in the desired proportions, heated to around 900$^\circ$C in air for 12 hours, then ground and sintered at 1450 $^\circ$C for 24 hours. The TEM samples used in the present study were prepared by crushing the well-characterized polycrystalline, and then the resultant suspensions were dispersed on a holey carbon-covered Cu grid. The crystallinity, shape and orientation of the different micrometer-sized flakes were checked by imaging and diffraction in our TEM (see Fig. 1 A, B, C).
Two samples were selected on the basis of the optimal shape and thickness for diffraction and EELS experiments. The sample for diffraction was 100 nm thick, while the one used for EELS was twice as thin (50 nm), to prevent severe multiple scattering effects \cite{multipsc}. The samples were then mounted on a double tilt holder and inserted in our fs-TEM, which is a JEOL 2100 modified for fs operation \cite{cptem} (the set-up is depicted in Fig. 1A). In femtosecond opeartion, statistically one electron per pulse was emitted at the gun. The repetition-rate of our experiments, both in diffraction and EELS, was 1 MHz resulting in an integrating time of 10 minutes per delay. In these conditions, the temporal resolution of our instrument was estimated to be 500 fs \cite{cptem}; because we did not observe sub-ps effects, we used 1 ps resolution in EELS and 2 ps resolution in diffraction to contain the integration time of the overall scan.
Fs electron and photoexciting pulses are generated by a Wyvern X Ti:Saph amplified laser capable of 5 W average power at rep-rates between 200 KHz and 2 MHz. The pulse duration was 80 fs and the photon energy was frequency doubled to obtain 400 nm linearly polarized pump pulses containing up to 1 $\mu$J of energy at 1 MHz repetition-rate. The pump beam was focused on the sample in a 100 $\mu$m spot, and the fluence was chosen to provide an average heating of the specimen sufficient to drive the material into its metallic phase, by suppressing the charge/orbital ordering that is naturally present at room temperature \cite{chao}. In Fig. 1, C,D,E,F and in the inset of Fig. 2 A, the diffraction patterns for the unperturbed sample, the sample excited by 5 mW at 1 MHz (corresponding to a fluence of 16 $\mu$J/cm$^2$), the sample excited by 20 mW , 40 mW, and 100 mW respectively are displayed. These static diffraction patterns were obtained with thermoionic electrons in order to see the effect of the average heating induced by the laser. The charge/orbital ordering satellites, indicated in the figure, are found to first rotate at intermediate laser fluence, corresponding to an average heating that brings the sample temperature from 300 K to about 320 K, while they have completely disappeared for an average pump power of 100 mW (320 $\mu$J/cm$^2$; the latter is the fluence used in our time-resolved experiments at which the sample is deeply in its metallic phase, at a temperature above 350 K). Upon irradiating the sample with the laser at 1 MHz repetition-rate, an increase in the sample temperature was observed and stabilized within few minutes, indicating that a new equilibrium was reached at a temperature higher than room temperature itself. In this condition, it was verified that the system relaxes to such a new thermal equilibrium between subsequent pulses, as testified by the flat temporal dependence of the diffraction intensity before time-zero, in Fig. 2. 

% Results

% diffraction

Snapshots of the crystal structure at different time-delays between the pump and the probe are obtained via fs electron diffraction. The sample is oriented in such a way that the electron beam impinges along the direction perpendicular to the $ab$-plane of the material. The diffraction patterns, in Fig. 1 C,D,E,F and Fig.2 inset, show the lattice parameters $a,b$ = 5.46 $\AA$, and the satellites at $q_a, q_b$ $\simeq$ 0.45. In the metallic phase (Fig. 2 inset), the satellites are absent. 
Light excitation is performed via 400 nm laser pulses, in the interband transitions region of the optical conductivity \cite{prmno1}. At positive time delays, the Bragg peak intensities are found to decrease, according to the behavior dictated by the Debye-Waller effect. The transient data along the in-plane diagonal of the unit cell (1,1,0) are depicted in Fig. 2. An exponential intensity drop of as much as 18$\%$ with a time constant of 31 ps is observed; this, according to the Debye formula: $ln(I/I_0) = -s^2<u^2>/3$ would correspond to an average atomic displacement induced by the temperature jump of around 3$\%$ of the interatomic distance. In Fig. 2 B, the change of the in-plane diagonal is shown; an expansion of 3$\%$ of the equilibrium value is indeed observed within the same time-scale (obtained via an exponential fit to be 31 ps). The distortion of the $c$-axis lattice parameter also contributes to the DW effect and is expected to be of the same size. In the dynamics of the Bragg peak intensity, an oscillation originating from the pressure-wave launched by the photoexcitation is visible. In this scenario, the sample behaves like the membrane of a drum, vibrating at particular frequencies that coincide with the film's eigenmodes \cite{nanodrum}; in our measurements, these oscillations have a period of 48 ps. It is possible to estimate the Young modulus ($Y$) along the $c$-axis knowing the thickness and the density of the material through the formula that relates $Y$ to the oscillation period of an ideal freely vibrating nanofilm: $\frac{1}{\tau_p}=\frac{n}{2d}(\frac{Y}{\rho})^{1/2}$, with $n$ = 1 for the fundamental tone. For the single crystal used for diffraction, assuming a density that for manganites typically is around 6.5g/cm$^3$ \cite{mndens}, we obtain a value of 113 GPa.

% EELS

In electron energy loss spectroscopy, multiple scattering effects complicate the assignment of the different spectroscopic features by inducing broad satellite peaks \cite{multipsc}. To avoid this problem, thinner samples must be used. For this reason, the mesoscopic sample we selected to perform ultrafast EELS was twice as thin (50 nm) the one observed in diffraction. The static spectra of our PrSr$_{0.2}$Ca$_{1.8}$Mn$_2$O$_7$ sample are displayed in Fig. 3 A in the range from 0 to 700 eV. Each region of this broad spectrum is modeled by theoretical calculations; in particular, in the low loss region, between 0 and 70 eV, two plasmon peaks at 3 and 13 eV are observed and are assigned to the partial and the highly-damped valence electron plasmon respectively. At higher energy, the shallower core levels are found, such as the Pr O$_{2,3}$ at 20 eV, the Sr N$_{2,3}$ at 29 eV, the Ca M$_{2,3}$ 34 eV and the Mn M$_{2,3}$ at 55 eV. This portion of the spectrum is modeled via state of the art DFT calculations via Wien2k code \cite{wien2k}, see red line in Fig. 3 A. The calculations are in reasonable agreement with the experiment, despite a shift of some of the electronic states already observed in previous reports \cite{lowloss}. At 345 eV, the Ca L$_{2,3}$ is observed and the 3 eV spin-orbit splitting is evident on this edge. In the oxygen K-edge instead, the sharp peak at 524.4 eV originates from the hybridized states between O-2$p$ and Mn-3$d$ orbitals, while two pronounced peaks centered at about 532 and 540 eV come from the hybridization of the O-2$p$ with Pr/Sr/Ca and Mn-4$s$/4$p$ states. The splitting between the valence band and upper Hubbard band is invisible here, because it is smaller than the energy resolution ($\simeq$ 1 eV in our fs-TEM). In the DFT calculations, red trace in Fig. 3 A, only the strong feature originating from the Pr-O hybridization at 533 eV is not captured. 
To simulate the Mn L-edge, the atomic multiplets code is used \cite{cowan}. The EELS spectrum is simulated by a mixed valence state Mn$^{3+}$ +  Mn$^{4+}$, in a crystal field environment of 2.4 eV as estimated by the Mn bands splitting in the high symmetry points of the fat bands diagram \cite{SI}. The broad experimental energy range provides a solid constraint to our modeling of the EELS spectra, which we use to understand the following time-resolved experiments.
Upon light excitation, as shown in diffraction, the film undergoes a periodic structural modulation as well as an overall dilatation due to the light induced temperature jump. The period of the modulation depends on the sample thickness, its Young modulus and density. For EELS experiments, the 50 nm thick sample used gives rise to a faster oscillation of the dynamical spectra with a period of 23 ps. The overall energy-time map of the experimental data is depicted in Fig. 3 B, where black lines indicate the energies corresponding to the different electronic states observed in the static spectra. Each of these states shows a different temporal dynamics, reflecting a different sensitivity to the photoinduced structural distortions. This is better visible in Fig. 3 C, where the temporal profile corresponding to each energy level is shown. Here, an overall decay of the peak intensity is observed after time zero, with a periodic modulation superimposed. It is interesting to notice that despite the overall effect is the weakest on the Mn M-edge at 53 eV, the periodicity is the clearest at this energy, suggesting that the Mn orbitals are the most sensitive to the structural distortions. This is not surprising considering the large crystal field on the Mn ions, which we estimated from the band structure and the Mn L-edge static spectrum, and that is certainly affected by the shape and dimension of the cage surrounding the Mn ions. 

% Discussion

To understand the dynamical EELS spectra, we performed DFT calculations of the low-loss as well as the oxygen K-edge region for different structural parameters, reflecting the changes estimated by diffraction, i.e. an overall lattice expansion by few percents of its equilibrium value. For comparison, we also calculate the spectrum for a compressed lattice, to verify the unambiguity of our interpretation. In Fig. 4 A, the difference spectra EELS(compr.-ab) - EELS(equil.), EELS(expand.-ab) - EELS(equil.), EELS(compr.-c) - EELS(equil.) and EELS(expand.-c) - EELS(equil.) are shown together with the experimental difference spectra EELS(t$<0$) - EELS(t=0) and EELS(t=50ps) - EELS(t=0). The sign of the transient change, i.e. a reduced EELS intensity, is captured by the calculations that consider an expanded lattice after time zero. The overall shape of the transient spectrum is also reasonably reproduced by the calculated difference spectra with a good coincidence between the position and relative strength of the negative peaks; the simulations consider a lattice expansion/compression in both $ab$ and $c$ direction of 2$\%$, and estimate intensity changes of the EELS features in the order of 3$\%$, and energy shifts in the order of 80-90 meV, close to what was observed in another layered solid, namely graphite \cite{carbonescience}. The magnitude of these changes is in good agreement with the experimental observation; the largest energy shift is observed at the Ca edge, causing a positive peak in the low energy side of the edge, in the theoretical curve. While the intensity change is captured by the experiment, the small energy shift (80 meV) is washed out by the overall energy resolution in time-resolved mode, around 2 eV. For this reason, the sharp positive peak is not visible in our transient data; the same situation was observed in previous time-resolved EELS data \cite{carbonecplth}.

In Fig. 4 B, the same simulation is performed for the oxygen K-edge spectrum. Interestingly, while in the low loss region of the spectrum (1-70 eV) an expansion of the in-plane or c-axis lattice parameters induces similar energy shifts and intensity changes of the different spectroscopic features, in the near-edge region of the oxygen K-edge, the behavior due to the in-plane or out of plane distortions is radically different. 
This is due to the fact that the O K-edge is very sensitive to the local Mn-O chemical bonding and consequently the orbital occupancy, while the low-loss spectra, including plasmons and other ionization edges, is mostly related to the bandwidth which depends on the volume expansion. In the charge/orbital ordered state, the orbital shape for Mn$^{3+}$ sites is dominated by the low-lying in-plane (3x$^2$-r$^2$) / (3y$^2$-r$^2$)-type orbitals which can be rotated through pressure or temperature changes. Therefore, changes in the lattices parameters along different directions would result in different orbital reconstruction. In addition, our calculations also demonstrate that the expanded $a$ and $b$ lattice parameters lead to a decrease in the energy gap, while the expanded $c$-axis lattice parameter has the opposite effect, as shown in the inset of Fig 4 B.
 
These results show that photoexcitation can be used to span the phase diagram of a manganite. In this particular case, we used light as a source of both continuous heating (to drive the system into its metallic phase) and to induce an ultrafast temperature jump in the material. The response to such a temperature jump was monitored via a combination of diffraction and EELS to observe the interplay between structural and electronic effects in the metallic phase of the sample. The only other ultrafst EELS experiment ever performed on a solid reported a radically different behavior in the first few ps of the dynamics for the layered system graphite \cite{carbonescience}. In that case, a compression of the system was observed before thermal expansion would take place. This behavior was attributed to the anisotropic decay of the out-of-equilibrium electronic structure, due to the peculiar semi-metallic band structure of graphite \cite{carbonescience}. In our results, a similar behavior was not observed within one ps resolution temporal scans. In EELS, a compression of the interlayer distances would give a large change of the plasmons, as estimated by our ab-initio caluclations (Fig. 4A, light blue trace) and because the out-of-plane bonding in layered solids is usually weak, soft (i.e. slow) phonons are involved in its distortions. For these reasons, a compression that is stronger and faster then predicted is unlikely (although not impossible); for these reasons, our data do not suggest that the behavior observed in graphite is general to all layered systems. Instead, a strong similarity between the layered manganite and graphite is observed in the long-time scale (several ps).
In both materials in fact, the light-induced thermal expansion is found to modulate the electronic structure in a similar way. In particular, the low-loss energy range of the EELS spectrum is shown to be able to discriminate between compression and expansion of the lattice, while being somewhat insensitive to the direction of the ionic motions, in-plane or out-of-plane. Based on these results, we also predict that distortions of the $ab$-plane or $c$-axis lattice parameters instead give very distinct spectroscopic signatures in the oxygen K-edge EEL spectrum.  

% Conclusions

These results provide a broad band direct observation of the interplay between the crystal and the electronic structure in a charge ordered manganite, and are the starting point for a deeper understanding of the different phase transitions in these materials. In fact, once the thermal effects are understood, it will be easier to distinguish dynamical electronic effects across charge ordering transitions. In particular, the fs dynamics of the orbital ordering rotation, as observed both via static conventional temperature dependent experiments, and the static photoinduced ones reported in Fig. 1 is currently under investigation by the same technique.

\begin{acknowledgments}
Work at LUMES was supported by the ERC starting grant USED258697, 
Work at Institute of Physics, CAS, was supported by National Basic Research Program of China 973 Program (Grant No. 2012CB821404) and Chinese Academy of Sciences (Grant No. YZ201258). work at BNL was supported by the U.S. DOE/BES, under Contract No. DE-AC02-98CH10886.
\end{acknowledgments}

%%%%%%%%%%%%%%%%%%%%%%%%%%%%%%%%%%%%%%%%%%%%%%%%%%%%%%%%%%%%%%%%

		\begin{figure*}[ht]
		\vspace*{.05in}
\begin{center}
\centerline{\includegraphics[width=190mm]{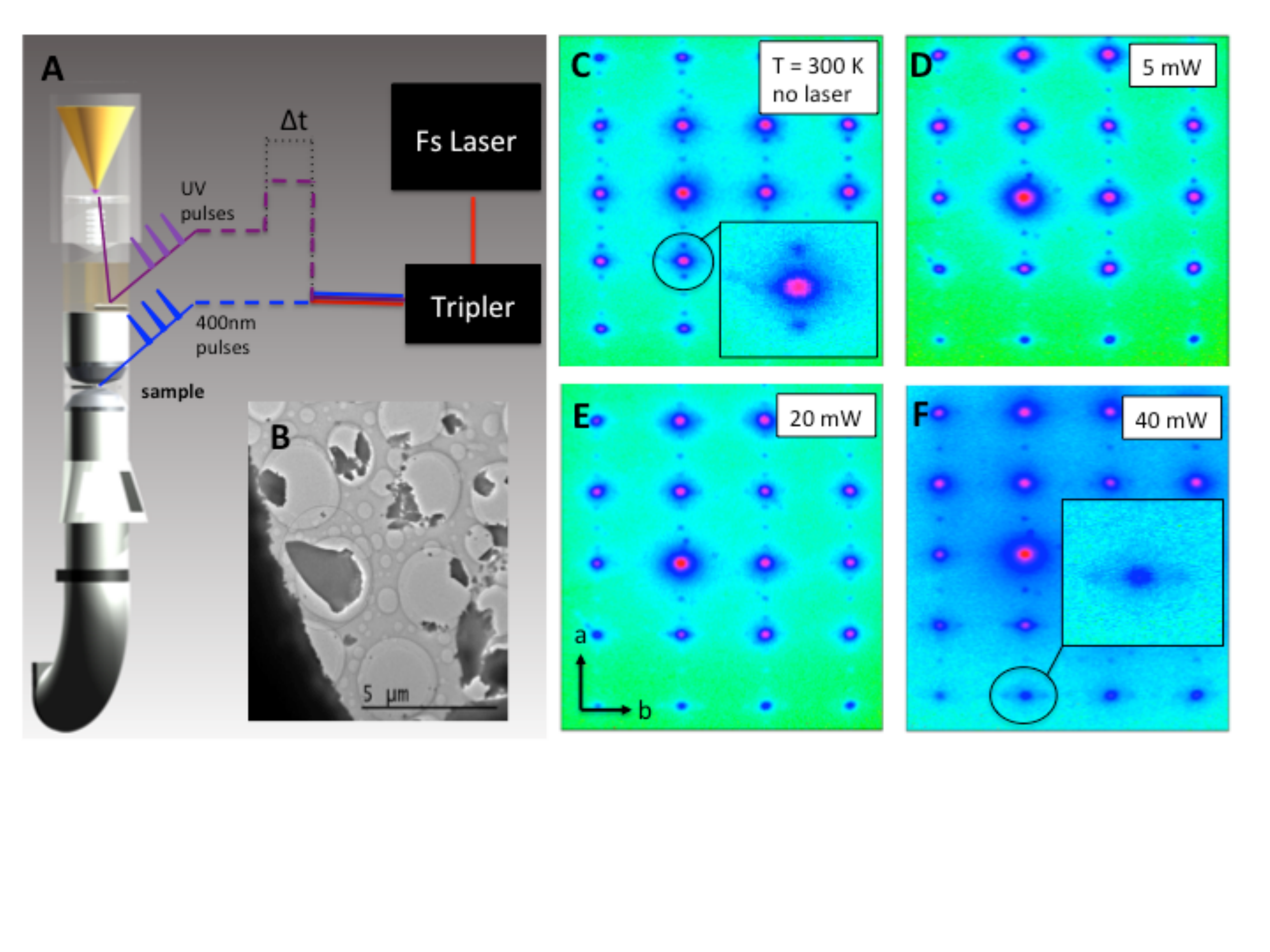}}
\caption{A Experimental set-up. The TEM column is depicted, the parts where the optical paths are found are transparent. B Image of the PrSr$_{0.2}$Ca$_{1.8}$Mn$_2$O$_7$ single crystalline nanoparticles. C Static unperturbed diffraction pattern at T=300 K. In the black circle the charge/orbital ordering satellites are evidenced. D, E, F static diffraction pattern for the sample irradiated by 5, 20 and 40 mW of laser light at 1 MHz respectively.}
\label{pseudo}
\end{center}
\end{figure*}
~\\
~\\
\newpage

		\begin{figure*}[ht]
		\vspace*{.05in}
\begin{center}
\centerline{\includegraphics[width=190mm]{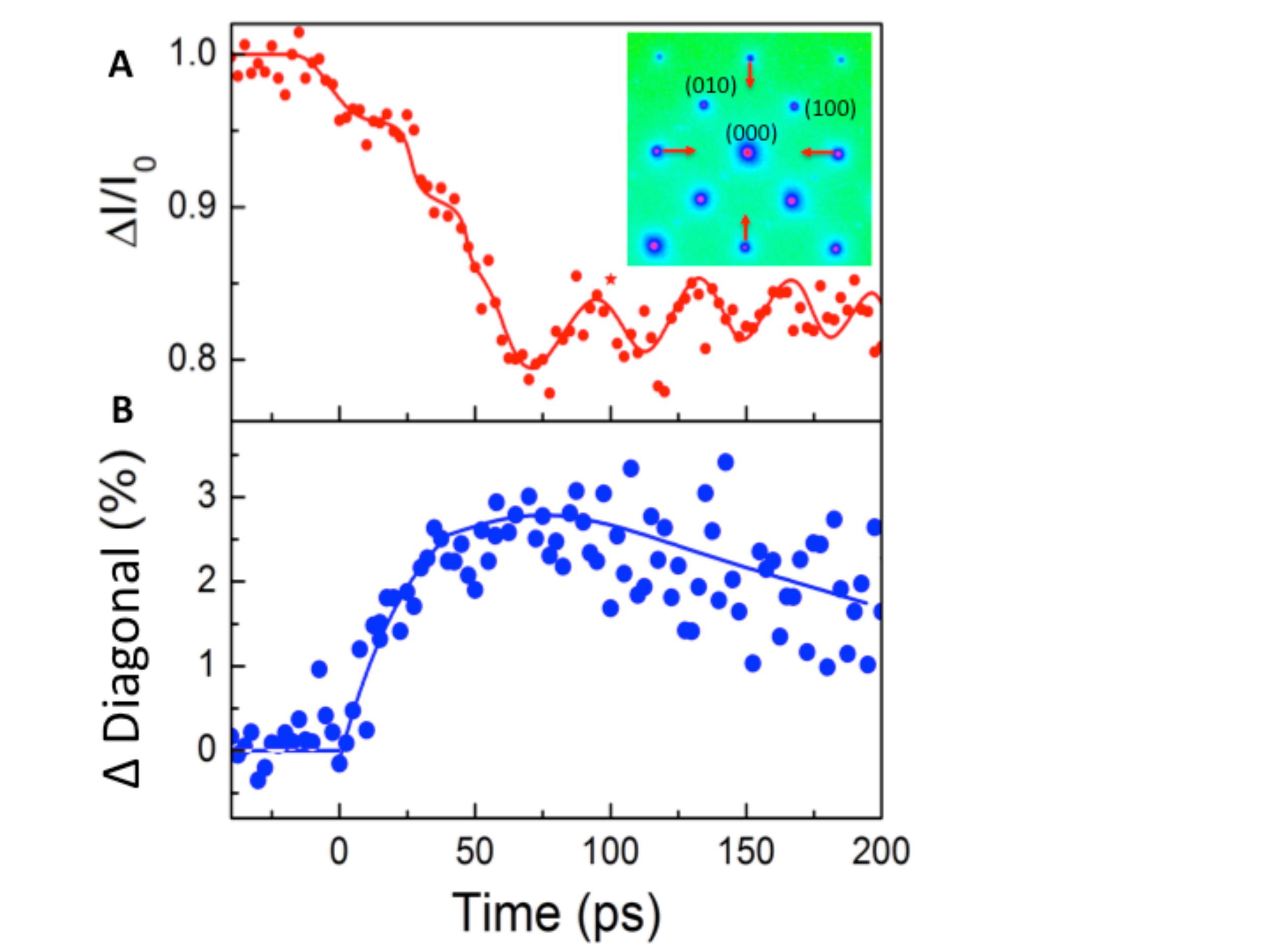}}
\caption{A Temporal evolution of the Bragg diffraction intensity. In the inset, the diffraction pattern for the sample irradiated by 100 mW of laser light at 1 MHz is shown. No charge/orbital ordering is evident in these conditions. B temporal evolution of the in-plane diagonal.}
\label{pseudo}
\end{center}
\end{figure*}
~\\
~\\
\newpage

		\begin{figure*}[ht]
		\vspace*{.05in}
\begin{center}
\centerline{\includegraphics[width=190mm]{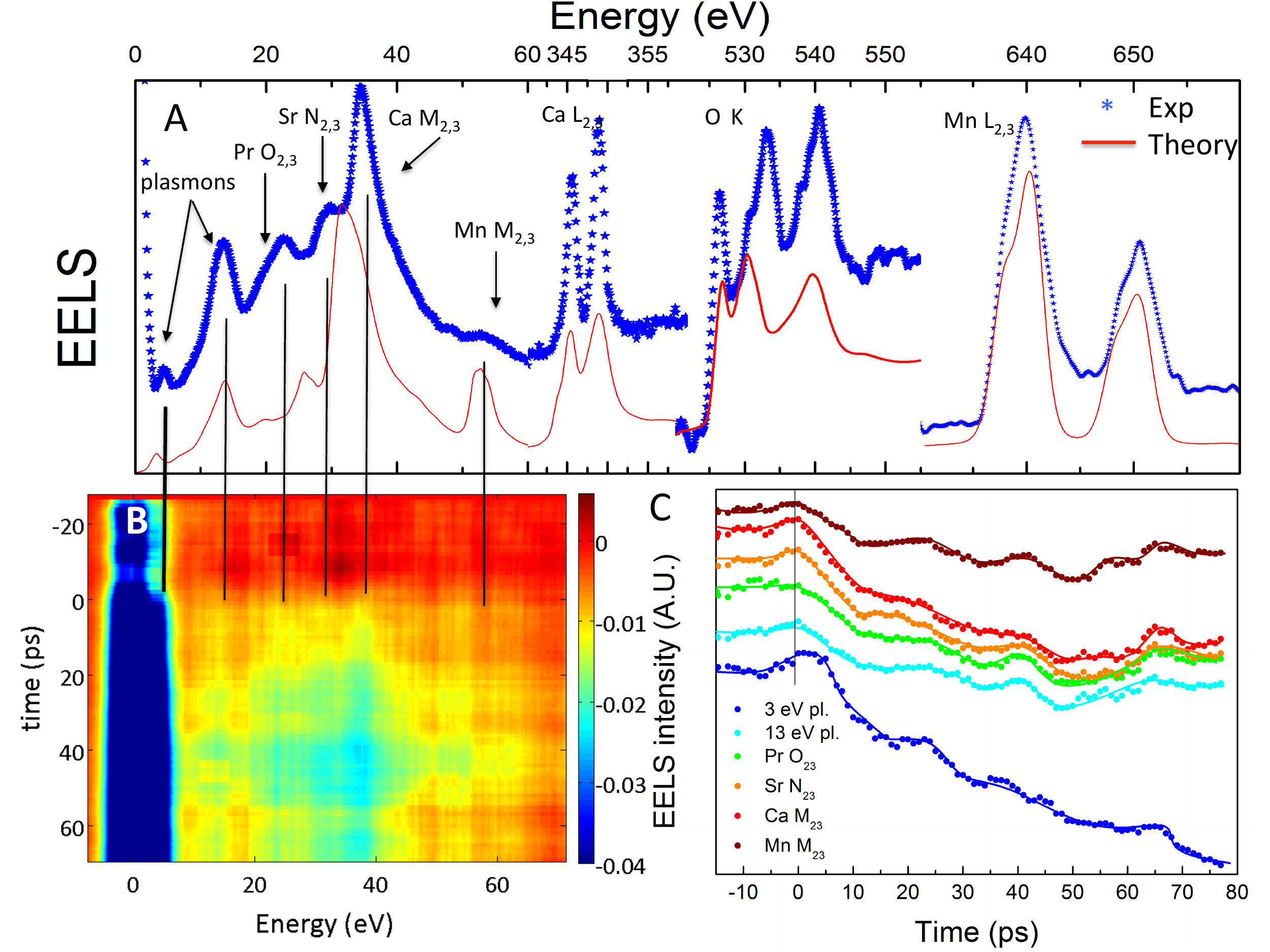}}
\caption{A Static EELS spectrum from 0 to 700 eV (blue symbols). The red curves are theoretical calculations; DFT caluclations are used until 600 eV and atomic multiplet calculations for the Mn L-edge above 600 eV. B Energy-time map of the low loss EELS spectrum. The 3D plot is obtained by taking the difference between EELS(t)-EELS(t$<0$). C Temporal profile of the EELS intensity at selected energies, in correspondence with the specified electronic states.}
\label{pseudo}
\end{center}
\end{figure*}
~\\
~\\
\newpage

		\begin{figure*}[ht]
		\vspace*{.05in}
\begin{center}
\centerline{\includegraphics[width=100mm]{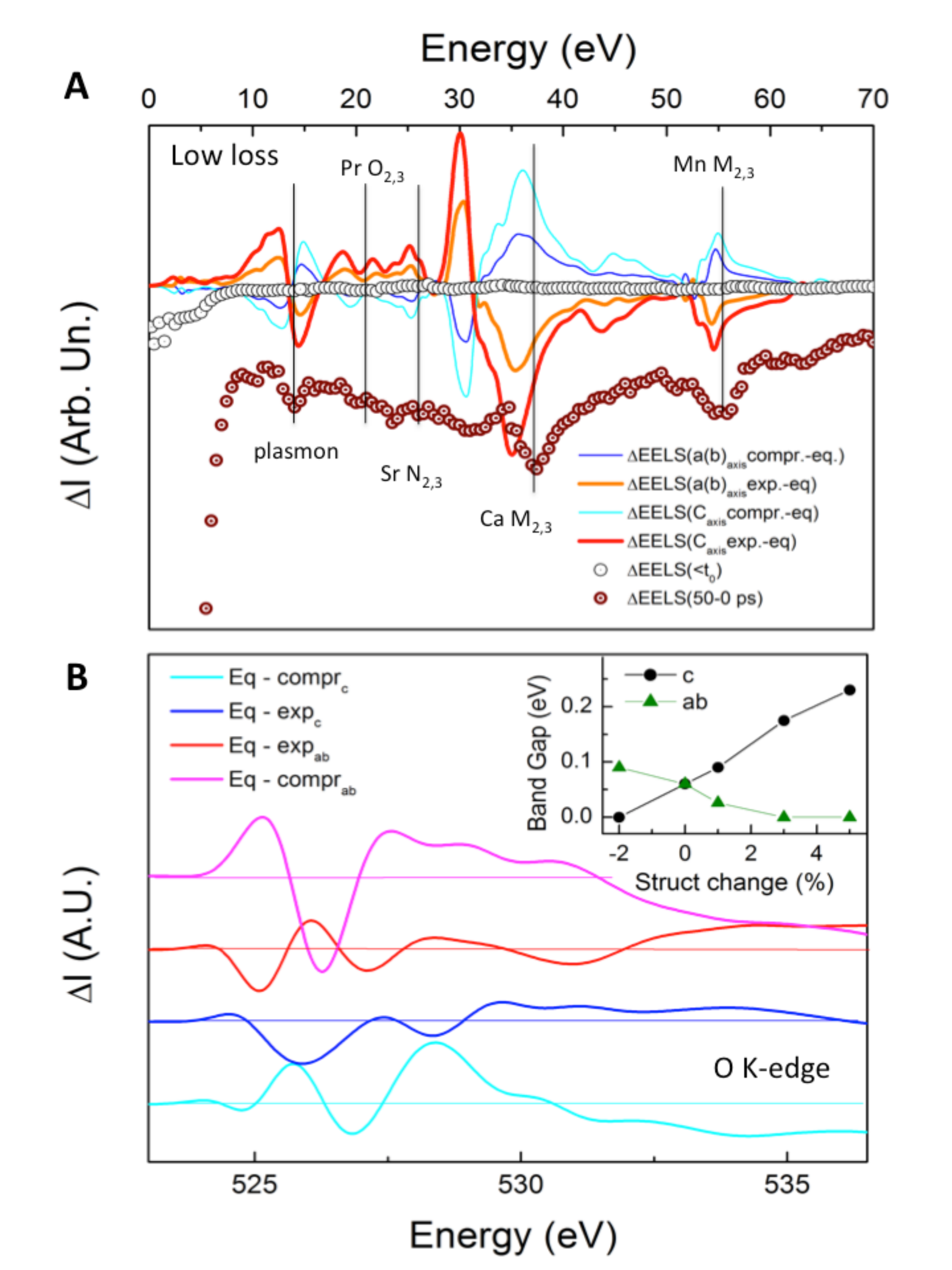}}
\caption{A The experimental differential EELS spectrum at t=50 ps is shown together with the calculated difference spectra for expanded/compressed $ab$-plane and $c$-axis. B Differential EELS spectra for the oxygen K-edge. The different lines are obtained by taking the difference between the K-edge spectrum at equilibrium and the spectrum for a compressed/expanded $a(b)$ or $c$-axis lattice parameters. In the inset, the evolution of the band gap as a function of the in-plane and out-of-plane lattice parameters change is shown. }
\label{pseudo}
\end{center}
\end{figure*}

~\\
~\\

\newpage
~\\

\section{Band structure}

In Fig. S1, the band structure of PrSr$_{0.2}$Ca$_{1.8}$Mn$_2$O$_7$ obtained by DFT is displayed.
The electronic structure calculations were performed by using the full potential linear augmented plane wave (LAPW) method within density functional theory (DFT) via the well-know WIEN2K package. The exchange correlation potential was treated by the generalized gradient approximation (GGA). The experimental lattice parameters and antiferromagnetic (AFM) structure were used. A supercell with Pr, Sr, and Ca atoms ordering was generated to treat the doping effect, according to the x-ray data. In order to treat the strong correlation effect of 3d and 4f electrons, different on-site Coulomb interactions (U) via GGA+U method were applied to Mn 3d (U = 3 eV) and Pr 4f (U = 5 eV) orbitals. The muffin-tin radii were set to 2.2, 2.1, 1.9, 2.0 and 1.5 for Pr, Sr, Ca, Mn and O atoms respectively. RKmax was set to 6 to determine the basis size. We used 16 irreducible Brillouin-zone k points for the generated supercell with 96 atoms. A fine k-mesh with more k points was generated for the simulations of the low-loss spectra and O-K edges which are calculated by the OPTIC and TELNES3 programs equipped in the WIEN2K package, respectively. All calculated spectra were broadened by 1 eV comparable with the energy resolution (about 1 eV) in the experimental spectra.

		\begin{figure*}[ht]
		\vspace*{.05in}
\begin{center}
\centerline{\includegraphics[width=100mm]{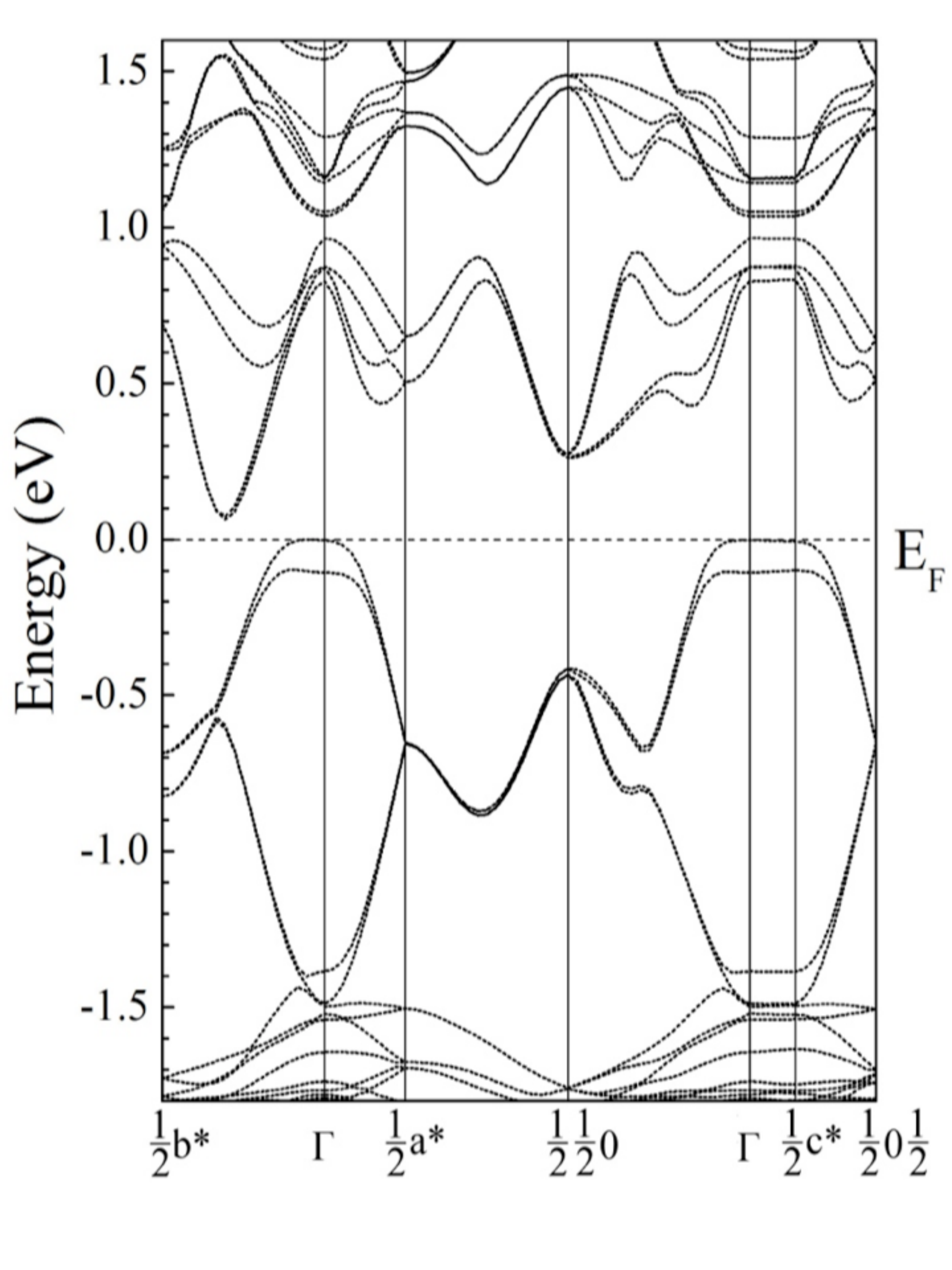}}
\caption{Fig. S1 Band structure of PrSr0.2Ca1.8Mn2O7 obtained by DFT, showing an insulating behavior (with a small energy gap of about 0.1 eV which is usually underestimated by the GGA+U method) in the initial state. The expansion along a(b) or c direction will results in a shift of the electronic states around the Fermi level which are dominated by the hybridized states between Mn-3d and O2p orbitals. That is why the observed periodicity of the Mn M-edge is the clearest.}
\label{pseudo}
\end{center}
\end{figure*}


\begin{thebibliography}{41}

\bibitem{graphenereview}
Novoselov, KS (2011) {\em Nobel Lecture: Graphene: Materials in the Flatland.} 
\textit{Rev. Mod. Phys.} 83:837.

\bibitem{manganitereview} 
Salamon, MB, Jaime, M (2001) {\em The physics of manganites: Structure and transport.}
\textit{Rev. Mod. Phys.} 73:583.

\bibitem{cuprates} 
Damascelli, A, Hussai, Z, Shen, ZX (2003) {\em Angle-resolved photoemission studies of the cuprate superconductors.}
\textit{Rev. Mod. Phys.} 75:473.

\bibitem{manganitesappl}
Haghiri-Gosnet, AM, Renard JP (2003) {\em CMR manganites: physics, thin films and devices.} 
\textit{J. Phys. D: Appl. Phys.} 36:R127.

\bibitem{manganitetemp}
Merz, M, et al. (2006) {\em Orbital degree of freedom in single-layered $La_{1-x}Sr_{1+x}MnO_4$: Doping- and temperature-dependent rearrangement of orbital states.}
\textit{Phys. Rev. B} 74:184414.

\bibitem{manganitepress}
Morimoto Y, Kuwahara, H, Tomioka, Y, Tokura, Y (1997) {\em Pressure effects on charge-ordering transitions in Perovskite manganites.}
\textit{Phys. Rev. B} 55:7549.

\bibitem{manganitemagfield}
Martin, C, Maignan, A, Hervieu, M, Raveau, B (1999) {\em Magnetic phase diagrams of $L_{1-x}A_xMnO_3$ manganites (L=Pr,Sm; A=Ca,Sr).}
\textit{Phys. Rev. B} 60:12191.

\bibitem{rini09}
Rini, M, et al. (2009) {\em Time-resolved studies of phase transition dynamics in strongly correlated manganites.}
\textit{J. of Phys.: Conference Series}  148:012013.

\bibitem{cav07nm}
Polli, D, et al. (2007) 
{\em Coherent orbital waves in the photo-induced insulator–metal dynamics of a magnetoresistive manganite.}
\textit{Nature Materials} 6:643.

\bibitem{cav07n}
Rini, M, et al. (2007)  {\em Control of the electronic phase of a manganite by mode-selective vibrational excitation.}
\textit{Nature} 449:72.

\bibitem{fskerr1} 
Miyasaka, K, Nakamura, M, Ogimoto, Y, Tamaru, H, Miyano, K (2006)
 {\em Ultrafast photoinduced magnetic moment in a charge-orbital-ordered antiferromagnetic $Nd_{0.5}Sr_{0.5}MnO_3$ thin film.} 
\textit{Phys. Rev. B} 74:012401.

\bibitem{fskerr2}
Ogasawara, T, et al. (2003)
 {\em Photoinduced spin dynamics in $La_ {0.6} Sr_ {0.4} MnO_ {3}$ observed by time-resolved magneto-optical Kerr spectroscopy.}
\textit{Phys. Rev. B} 68:180407(R).

\bibitem{taylor01}
Averitt, AD, et al. (2001)
{\em Ultrafast Conductivity Dynamics in Colossal Magnetoresistance Manganites.}
\textit{Phys. Rev. Lett.} 87:017401.

\bibitem{fskerr3}
Matsubara, M, et al. (2007) 
{\em Ultrafast Photoinduced Insulator-Ferromagnet Transition in the Perovskite Manganite $Gd_{0.55}Sr_{0.45}MnO_3$.}
\textit{Phys. Rev. Lett.} 99:207401.

\bibitem{cav11}
F\"orst, M, et al. (2011)
{\em Driving magnetic order in a manganite by ultrafast lattice excitation.}
\textit{Phys. Rev. B} 84:241104(R).

\bibitem{cav11b}
Ehrke, H et al. (2011)
{\em Photoinduced Melting of Antiferromagnetic Order in $La_{0.5}Sr_{1.5}MnO_4$ Measured Using Ultrafast Resonant Soft X-Ray Diffraction.}
\textit{Phys. Rev. Lett.} 106:217401.

\bibitem{steve1}
Beaud, P, et al. (2009)
{\em Ultrafast Structural Phase Transition Driven by Photoinduced Melting of Charge and Orbital Order.}
\textit{Phys. Rev. Lett.} 103:155702.

\bibitem{steve2}
Caviezel, A, et al. (2012)
{\em Femtosecond dynamics of the structural transition in mixed valence manganites.}
\textit{Phys. Rev. B} 86:174105.

\bibitem{fsxray}
Lee, HJ, et al. (2008)
{\em Optically induced lattice dynamics probed with ultrafast x-ray diffraction.} 
\textit{Phys. Rev. B} 77:132301.

\bibitem{ichi11}
Ichikawa, I, et al. (2011)
{\em Transient photoinduced ‘hidden’ phase in a manganite.}
\textit{Nature Materials} 10:101.

\bibitem{chao}
Li, ZA, et al. (2009)
{\em A “checkerboard” orbital-stripe phase and charge ordering transitions in $Pr(Sr_xCa_{2−x})Mn_2O_7$ (0 $<$ x $<$ 0.45).}
\textit{Europhys. Lett.} 86:67010.

\bibitem{jure}
Eichberger, M, et al. (2010)
{\em Snapshots of cooperative atomic motions in the optical suppression of charge density waves.}
\textit{Nature} 468:799.

\bibitem{baum}
Baum, P, Yang, DS, Zewail, AH (207)
{\em 4D Visualization of Transitional Structures in Phase Transformations by Electron Diffraction.}
\textit{Science} 318:788.

\bibitem{nuh}
Gedik, N, Yang, DS, Logvenov, G, Bozovic, I, Zewail, AH (2007)
{\em Nonequilibrium Phase Transitions in Cuprates Observed by Ultrafast Electron Crystallography.}
\textit{Nature} 316:425.

\bibitem{carbonescience}
Carbone, F, Kwon, OH, Zewail, AH (2009)
{\em Dynamics of Chemical Bonding Mapped by Energy-Resolved 4D Electron Microscopy.}
\textit{Science} 325:181.

\bibitem{carbonecpl}
Carbone, F, et al. (2009)
{\em EELS femtosecond resolved in 4D ultrafast electron microscopy.}
\textit{Chem. Phys. Lett.} 468:107.

\bibitem{carboneperspective}
Carbone, F, Musumeci, P, Luiten, OJ, Hebert, C (2012)
{\em A perspective on novel sources of ultrashort electron and X-ray pulses.}
\textit{Chem. Phys.} 392:1.

\bibitem{prmno1}
Tokunaga, Y, et al. (2006)
{\em Rotation of orbital stripes and the consequent charge-polarized state in bilayer manganites.}
\textit{Nature Materials} 5:937.

\bibitem{phased}
Li, QA, et al. (2007)
{\em Reentrant Orbital Order and the True Ground State of $LaSr_2Mn_2O_7$.}
\textit{Phys. Rev. Lett.} 98:167201.

\bibitem{brettscience}
Barwick, B, Park, HS, Kwon, OH, Baskin, JS, Zewail, AH (2008)
{\em 4D Imaging of Transient Structures and Morphologies in Ultrafast Electron Microscopy.}
\textit{Science } 322:1227.

\bibitem{nanodrum}
Kwon, OH, Barwick, B, Park, HS, Baskin, JS, Zewail, AH (2008)
{\em Nanoscale Mechanical Drumming Visualized by 4D Electron Microscopy.}
\textit{Nano Lett.} 8:3557.

\bibitem{multipsc}
Egerton, RF (2009)
{\em Electron Energy-Loss Spectroscopy in theTEM.}
\textit{Rep. Prog. Phys.} 72:016502.

\bibitem{cptem}
Piazza, L, et al. (2013)
{\em Design and implementation of a fs-resolved transmission electron microscope based on thermionic gun technology.}
\textit{Chem. Phys.} 423:79. 

\bibitem{mndens}
Armstrong, TJ, Virkar, AV (2002)
{\em Performance of Solid Oxide Fuel Cells with LSGM-LSM Composite Cathodes.}
\textit{J. Electrochem. Soc.} 149:A1565.

\bibitem{wien2k} Information on this software can be found at: www.wien2k.at.

\bibitem{lowloss}
Keast, VJ (2007)
{\em Ab initio calculations of plasmons and interband transitions in the low-loss electron energy-loss spectrum.}
\textit{J. Electr. Spectr. Rel. Phenom.} 143:97.

\bibitem{cowan} Information on this software can be found at: www.tcd.ie/Physics/people/Cormac.McGuinness/Cowan/.

\bibitem{SI} See supplementary material at [URL will be inserted by AIP] for the details of the electronic structure calculations and the corresponding band diagram.

\bibitem{carbonecplth}
Carbone, F (2010)
{\em The interplay between structure and orbitals in the chemical bonding of graphite.}
\textit{Chem. Phys. Lett.} 496:291.

\end{thebibliography}
\end{document}